\documentclass[letterpaper, 10 pt, conference]{ieeeconf}  

\IEEEoverridecommandlockouts                              

\usepackage{graphics} 
\usepackage{epsfig} 
\usepackage{mathptmx} 
\usepackage{times} 
\usepackage{amsmath} 
\usepackage{amssymb}  
\usepackage{multirow}
\usepackage{url}
\usepackage{arydshln}

\usepackage{cite}

\title{\LARGE \bf
Differential algebraic modeling of an alkaline electrolyzer plant
}

\author{Nicola Cantisani, Josefine Dovits, John Bagterp J{\o}rgensen
\thanks{N. Cantisani, and J.B. Jørgensen are with Department of Applied Mathematics and Computer Science, Technical University of Denmark, DK-2800 Kgs. Lyngby, Denmark}%
}

\begin{document}

\maketitle
\thispagestyle{empty}
\pagestyle{empty}

\begin{abstract}
We develop a mathematical model for dynamic simulation of an alkaline electrolyzer plant. The plant includes the stack, a water recirculation system and hydrogen storage with compressor. We model each component of the system with mass and energy balances. Our modeling strategy consists of a rigorous and systematic formulation using differential algebraic equations (DAE), along with a thermodynamic library that evaluates thermophysical properties. We perform a simulation with step power input. Dynamic modeling enables simulation and model-based optimization and control for optimal hydrogen production under varying operating conditions.
\end{abstract}

\section{Introduction} \label{sec:introduction}
In the fight against climate change, renewable energy sources and electrification are undoubtedly regarded as the main characters, and research and development are pushing towards a rapid expansion in that sense. However, electrification does not seem to be a viable option for all kind of energy-intensive activities, like, for example, heavy transport (ships or aircrafts). Alternative green fuels shall instead be considered. The process of converting renewable energy into green fuels is called Power-2-X. Amongst these, hydrogen has a central role. Hydrogen is considered a good option because of its high energy density and because it does not produce any pollutant gas when used \cite{OLIVIER2017280} \cite{DAVID2019392}. Moreover, hydrogen is used to produce ammonia, which can be used as a fuel or a fertilizer.
Water electrolysis is the chemical process of splitting water into hydrogen and oxygen by the use of electrical current. The hydrogen is defined green if electrolysis is performed using renewable energy. 
Because of their nature, renewable energy sources (e.g. wind and solar) are intrinsically stochastic and not controllable. Moreover, as the share of renewables in the electric grids grows, extra flexibility is needed to overcome power imbalances. Electrolyzers can create considerable extra economic revenue by functioning as ancillary services, i.e. frequency containment reserve \cite{Samani2020} \cite{johnsen2023value}. In this new energetic scenario, electrolyzers are intended to operate dynamically, rather than at steady state. 
Therefore, dynamic modeling of the electrolysis process is necessary to simulate the effect of using intermittent varying electric power.


There is extensive literature on modeling an alkaline electrolyzer stack. The main electrochemical model goes back to \cite{ULLEBERG200321}. More recent contributions have also considered the mass and energy balances in more detail, for example \cite{SAKAS20224328} \cite{RIZWAN202137120}. \cite{OLIVIER2017280} provides a review of the relevant literature.
There is also research on how to control and optimize the operation of an electrolyzer, for example \cite{HUANG202316} \cite{QI2023120551}.

This paper deals with formulating a dynamic model of an alkaline electrolyzer plant. The plant consists of an electrolyzer stack (the core element) and its peripherals: two liquid-gas separators (one for hydrogen and one for oxygen), a compressor, and a hydrogen storage tank. The plant also includes a water recirculation system with a flow mixer and heat exchangers. We model each component of the system in a rigorous and systematic way. Our modeling approach is based on physical first principles, i.e. mass and energy balances. The novelty of the paper consists in mainly two features of the model. The former is the rigorous formulation using DAEs. The latter is the use of a thermodynamic library to precisely evaluate thermophysical properties of gas and liquids. The DAE formulation makes the model easy to implement, thanks to the rich literature on simulation (and also control) of DAEs. The use of a thermodynamic library makes the formulation more accurate (and more general), as opposed to assuming constant heat capacity of the materials.
The model can ultimately be used as a proxy for control and optimization. This means that, for example, it can be used for performing simulations for tuning or evaluating the performance of a controller.
The effect of each component on the overall performance can be easily assessed.

The paper is structured as follows. Section \ref{sec:overview} provides an overview and a description of the water electrolysis process, focusing on the alkaline technology. Section \ref{sec:dynmodelling} presents our formulation of the methods used for the formulation of dynamic models, based on physical laws. Section \ref{sec:model} deals with modeling all the components of the system using DAEs. Section \ref{sec:simulation} presents some simulation results of the full model. 
Section \ref{sec:conclusion} concludes the paper.

\begin{figure*}[tb]
    \centering
    \smallskip
    \includegraphics[width=\textwidth]{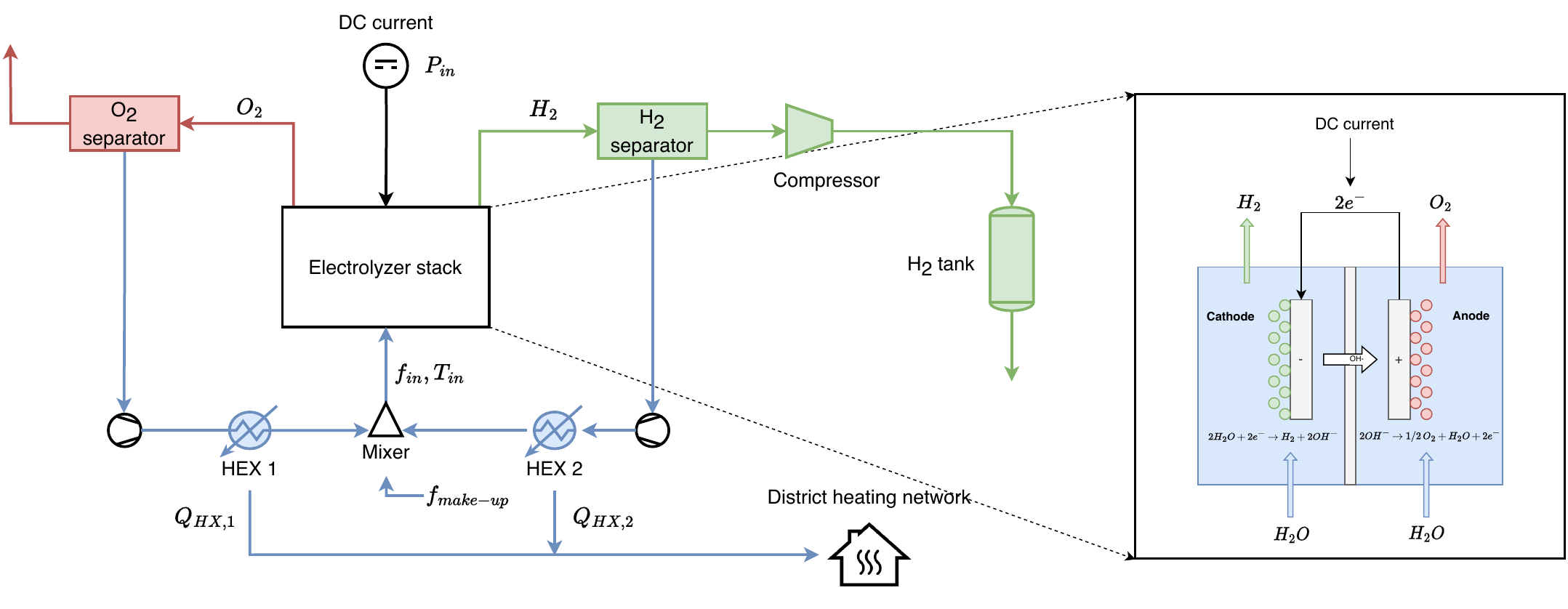}
    \caption{On the left: process sketch of the electrolyzer plant. On the right: working principle of an alkaline electrolytic cell. The reduction reaction happens at the cathode, where hydrogen is produced. The oxidation reaction happens at the anode, where oxygen is produced.}
    \label{fig:process_sketch}
\end{figure*}
\section{Overview of water electrolysis} 
\label{sec:overview}
\subsection{Water electrolysis}
The water splitting reaction is
\begin{equation}
    \mathrm{H_2O (l) + \text{Electrical power} \longrightarrow H_2 (g) + \frac{1}{2} O_2 (g)}.
\end{equation}
There are different technologies available to perform the reaction. We consider alkaline electrolysis. It is characterized by having two electrodes (cathode and anode), operating in a liquid alkaline electrolyte solution of potassium hydroxide (KOH), usually 20-40\%, or sodium hydroxide (NaOH), separated by a diaphragm. It operates at relatively low temperatures (max $90 {}^\circ C$). The working principle is depicted in Fig. \ref{fig:process_sketch}. The half reactions are
\begin{align}
   \text{Cathode (red.): } & \mathrm{ 2 H_2O(l) + 2e^-	\longrightarrow	H_2(g) + 2 OH^-(aq), }\\
   \text{Anode (ox.): } &\mathrm{  2 OH^-(aq) \longrightarrow 1/2 O_2(g) + H_2O(l) + 2 e^-.}
\end{align}
\subsection{Process description}
We consider the electrolyzer plant sketched in Fig. \ref{fig:process_sketch}. 
The core of the process is the electrolyzer stack. 
It receives a flow of water and electric power. 
The produced hydrogen and oxygen gas leave the stack with the water stream, from their respective chambers. 
The two flows reach a respective gas/liquid separator, where the gas phase is separated from the liquid phase due to the different densities. 
The water outflows of the separators are recirculated back into the electrolyzer.
The recirculated flows pass through heat exchangers, to remove excess heat produced in the electrolyzer, before they are mixed with fresh make-up water. 
The combined stream then enters the electrolyzer.
The heat removed in the two heat exchangers is, ideally, recycled by sending it to the district heating network.
The hydrogen gas leaving the hydrogen separator is compressed in a multistage compressor, to be stored under pressure in a hydrogen tank.
From the storage tank the hydrogen can be distributed to downstream processes or other facilities, for example.

\section{Dynamic modeling} \label{sec:dynmodelling}
We investigate models for water electrolysis in the form of differential algebraic equations (DAE)
\begin{subequations}
\begin{alignat}{3}
    \dot x &= f(x,y,u,d,p), \\
    0 &= g(x,y,u,d,p).
\end{alignat}
\end{subequations}
$x$ is the vector of differential variables, $y$ are algebraic variables, $u$ are manipulated variables (MVs), $d$ are the disturbances and $p$ is the vector of model parameters.

Additionally, the outputs of interest of the system may be denoted with
\begin{equation}
    z = h(x,y,u,d,p).
\end{equation}

\subsection{Modeling principles}
Our modeling approach is based on physical first principles. We use mass and energy balances to model each component of the system. We review them here briefly, in order to present our formulation and notation.

We express the change in time of the number of moles in a physical system as
\begin{equation}
    \dot n = f_{in} - f_{out} + R,
\end{equation}
where $f_{in}$ and $f_{out}$ are the molar inlet and outlet mass flow rates and $R$ is the production rate. Notice that we always assume vector notation for handling different chemical species. The production rate is given by the stoichiometry and kinetics of the chemical reaction happening in the system, i.e.
\begin{equation}
    R = S' r, \qquad r = r(T,P,n).
\end{equation}
$S$ is the stoichiometric matrix, $r$ is the reaction rate, $T$ is the temperature and $P$ is the pressure.
 
The total energy in a system is the sum of the potential, kinetic and internal energy
\begin{equation}
    E = E_p + E_k + U.
\end{equation}
We consider systems for which $E \approx U$ holds. The energy balance for such a system is
\begin{equation}
    \dot U = \tilde H_{in} - \tilde H_{out} + Q + W_s + W_e.
\end{equation}
We indicate with $\tilde H$ any enthalpy flow. $Q$ is any heat transferred to/from the external ambient, $W_s$ is the shaft work and $W_e$ is the electrical work.

\subsection{Thermodynamic functions}
We rely on a thermodynamic library to evaluate all the thermodynamic functions. We use ThermoLib \cite{RITSCHEL20173542}, a library implemented in Matlab and C that uses cubic equations of state to compute vapor and liquid phase thermodynamic properties. The library can also evaluate first and second order derivatives of the thermodynamic functions, which is often required when constructing the Jacobian of such dynamic models. ThermoLib evaluates the enthalpy, the entrophy and the volume as a function of the temperature $T$, pressure $P$ and number of moles $n$, i.e.,
\begin{subequations}
\begin{align}
H &= H(T,P,n), \\
S &= S(T,P,n), \\
V &= V(T,P,n).
\end{align}
\end{subequations}
Other thermodynamic properties can be easily derived.
\begin{subequations}
\begin{align}
U &= H - P V, \\
G &= H - T S.
\end{align}
\end{subequations}
$U$ is the internal energy, and $G$ is Gibbs' free energy.
Enthalpy flow rates are computed as
\begin{equation}
    \tilde H = H(T,P,f),
\end{equation}
where $f$ is a mass flow rate.

\section{Process modeling} \label{sec:model}
This section deals with modeling the main components of an electrolyzer plant, as depicted in Fig. \ref{fig:process_sketch}. We provide a description of each part and sum up the full model as a system of DAEs.

\subsection{Electrolyzer stack}
The electrolyzer stack consists of multiple electrolytic cells connected in series. The general working principle of an electrolytic cell is shown is Fig. \ref{fig:process_sketch}, to the right.

The model consists of an electrochemical part, and mass and energy balance.

\subsubsection{Electrochemical reaction}
The minimum voltage requirement for reaction, called the reversible voltage, is computed by
\begin{equation}
    \xi_{rev} = \frac{\Delta G_r}{z_e F}.
\end{equation}
$\Delta G_r$ is the Gibbs' free energy at the reaction conditions.
$F$ is the Faraday constant and  $z_e=2$ is the number of electrons transferred.
$\Delta G_r$ is calculated based on the pure component enthalpies $H_i$ and entropies $S_i$ at given temperature $T$, operating pressure $P$ for liquid species and partial pressure $p_i$ for the gas species.
\begin{subequations}
    \begin{align}
        \Delta H_r & = H_{H_2}(T,p_{H_2})(g) + \frac{1}{2} H_{O_2}(T,p_{O_2})(g) - H_{H_2O}(T,P)(l), \\
        \Delta S_r & = S_{H_2}(T,p_{H_2})(g) + \frac{1}{2} S_{O_2}(T,p_{O_2})(g) - S_{H_2O}(T,P)(l), \\
        \Delta G_r & = \Delta H_r - T \Delta S_r.
    \end{align}
\end{subequations}
(l) or (g) indicate liquid or gas species.
The pressure and temperature dependence of the reversible voltage is considered by evaluating $H_i$ and $S_i$ at appropriate temperature and pressure.
We assume that the gas phase in the cathode and the anode chamber is pure hydrogen and pure oxygen, respectively, therefore $p_{H2} = p_{O2} = P$.
The enthalpy and entropy of water is calculated at operating pressure $P$, assuming that the presence of KOH in the solution can be neglected and water therefore behaves like an ideal liquid.
We assume that the pressure is constant.
ThermoLib is used to evaluate the $H_i$ and $S_i$, which has the added benefit that nonlinearities of the heat capacities, as modelled in the DIPPR (Design Institute for Physical Properties) database, are taken into account.

The total energy requirement is expressed by the reaction enthalphy
\begin{equation}\label{eq:enthalphy}
    \Delta H_r = \Delta G_r + T \Delta S_r.
\end{equation}
$T$ is the temperature and $\Delta S_r$ is the reaction entropy.
The second term in \eqref{eq:enthalphy} accounts for the irreversibilities of the reaction. This extra energy requirement is therefore brought as heat.

The \textit{real} total voltage of a single electrolytic cell takes into account extra irreversibilities \cite{ULLEBERG200321}
\begin{equation}
\xi_{cell} = \xi_{rev} + \xi_{ohm} + \xi_{act} + \xi_{con}.
\end{equation}
$\xi_{ohm}$ is the ohmic overvoltage. This is due to ohmic losses in the cell, generated by the resistance to the electron flux in the electrodes, the connectors and also by the resistance in the electrolyte and the diaphragm between the 2 chambers. $\xi_{act}$ is the activation overvoltage. It comes from the kinetics of the electric charge transfer that occurs at the electrode reaction sites. This in other words describes the activation barrier in the electrochemical reaction. $\xi_{con}$ is the concentration overvoltage. This is caused by mass transfer losses, and it happens only at very high current densities, hence it can be neglected for alkaline electrolyzers.

The following empirical model for the activation and ohmic overvoltages has been proposed by \cite{ULLEBERG200321}
\begin{subequations} 
    \begin{align}
    \xi_{ohm} & = (r_1 + r_2 T)\frac{I}{A}, \\
    \xi_{act} & = s \log \left( \left(t_1 + \frac{t_2}{T} + \frac{t_3}{T^2} \right) \frac{I}{A} + 1 \right).
    \end{align}
\end{subequations}
$A$ is the electrode area of a single cell, $I$ is the current. The parameters $r_1$ and $r_2$ represent the ohmic resistance, while $s$ and $t_1,t_2,t_3$ are the activation overvoltage coefficients. These are estimated empirically and depend on the specific electrolyzer. The voltage and the current are related to the input electric power 
\begin{equation}
 P_{in} = n_c \xi_{cell} I.
\end{equation}
$n_c$ is the number of (identical) electrolytic cells connected in series.
The reaction rate is a function of the current
\begin{equation}
    r(I) = n_c \frac{\eta_F}{zF} I.
\end{equation}
The term $\eta_F$ is known as Faraday efficiency. This is 
\begin{equation}
    \eta_F = \frac{(I/A)^2}{f_1 + (I/A)^2} f_2.
\end{equation}
$f_1$ and $f_2$ are parameters.

The electrochemical dynamics enter the model as algebraic equations
\begin{equation}
    g_{el} = \begin{bmatrix}
        \xi_{cell} - \xi_{rev}(T) - \xi_{ohm}(I,T) - \xi_{act}(I,T) \\
        P_{in} - n_c \xi_{cell} I
    \end{bmatrix} = 0.
\end{equation}
    
\subsubsection{Mass balance}
We compute the outgoing flow streams from the anodic and cathodic chambers assuming steady state in the mass balance. We use vector notation with chemical species in the order [H$_2$O, H$_2$, O$_2$]$^T$, i.e.
\begin{equation}
    n = \begin{bmatrix}
        n_{H_2O} \\ n_{H_2} \\ n_{O_2}
    \end{bmatrix}.
\end{equation} 
In the anodic chamber we have
\begin{equation}
    \dot n^{an} = \frac{1}{2} f_{in} - f_{out}^{an} + R^{an} = 0.
\end{equation}
The outgoing flow is 
\begin{equation}
    f_{out}^{an} = \begin{bmatrix}
    f_{in, H_2 O}/2 \\ 0 \\ 0
    \end{bmatrix} + \begin{bmatrix}
        1 \\ 0 \\ 1/2
    \end{bmatrix} r.
\end{equation}
In the chatodic chamber we have
\begin{equation}
    \dot n^{cat} = \frac{1}{2} f_{in} - f_{out}^{cat} + R^{cat} = 0.
\end{equation}
The outgoing flow is 
\begin{equation}
    f_{out}^{cat} = \begin{bmatrix}
    f_{in,H_2 O}/2 \\ 0 \\ 0
    \end{bmatrix} + \begin{bmatrix}
        -2 \\ 1 \\ 0
    \end{bmatrix} r.
\end{equation}
Notice that it is assumed that the inlet flow of water is split exactly in equal parts between the anode and cathode chambers.
The total outgoing flow from the stack is
\begin{equation}
    f_{out} = f_{out}^{an} + f_{out}^{cat}.
\end{equation}

\subsubsection{Energy balance}
The energy balance in the electrolyzer is
     \begin{equation}
    \dot U_{el} = \tilde{H}_{in} - \tilde{H}_{out} + Q + W_e.
\end{equation}
$U_{el}$ is the internal energy of the electrolyzer, $\tilde{H}_{in}$ and $\tilde{H}_{out}$ are the enthalpy flows in and out, $Q$ is heat flow exchanged with the ambient (convective) and $W_e$ is the electrical work. Specifically
\begin{subequations}
    \begin{align}
    W_e  =& P_{in}, \\
    Q  =& - Q_{amb} = - A_s h_c ( T - T_{amb} ),\\
    \tilde{H}_{in}  =& H(T_{in},P,f_{in}) = H(T_{in},P,f_{in,H_2O})(l), \\
    \tilde{H}_{out}  =& H(T,P,f_{out}) =  H(T,P,f_{out,H_2O})(l),  \\ & + H(T,P,f_{out,H_2})(g) + H(T,P,f_{out,O_2})(g). \nonumber
    \end{align}
\end{subequations}
$A_c$ is the active area of heat transfer, $h_c$ is the heat transfer coefficient and $T_{amb}$ is the ambient temperature.
The internal energy of the electrolyzer is a function of the form
\begin{equation}
    U_{el} =  U_{el}^0 + C_{p,el} (T - T^0),
\end{equation}
with 
\begin{equation}
    C_{P,el} = C_{P,H_2O} + C_{P,H_2} + C_{P,O_2} + C_{P,metal}.
\end{equation}
We assume that the mass of fluids and the metal are constant, therefore $C_{P,el}$ can be considered constant. As a consequence, $\dot U_{el} = C_{P,el} \dot T$, and $T$ is used as state variable in practice.





\subsection{Liquid/gas separator}
\begin{figure}[tb]
    \centering
    \smallskip
    \includegraphics[width=0.27\textwidth]{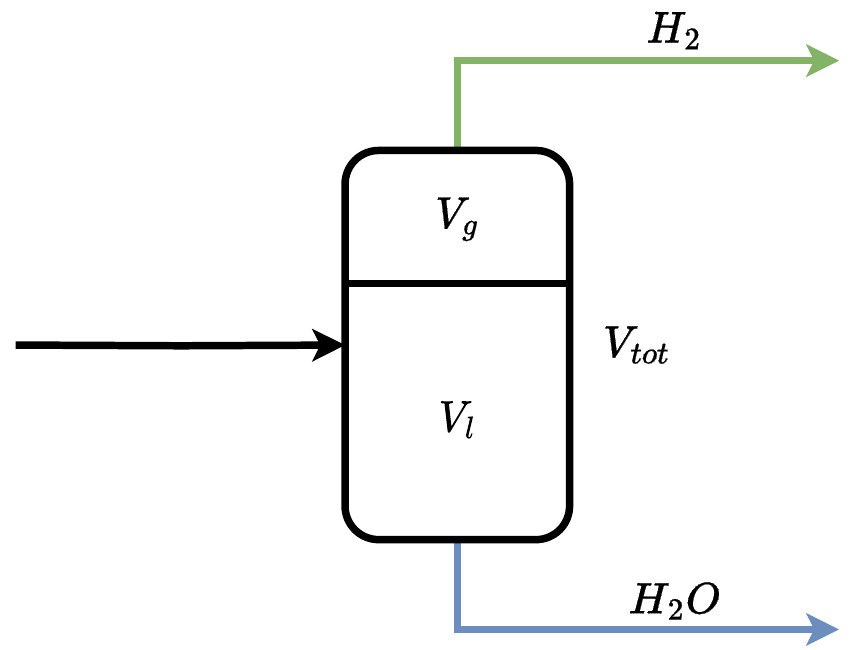}
    \caption{Gas-liquid separator for the hydrogen stream out of the electrolyzer.}
    \label{fig:separator}
\end{figure}

We model the liquid and gas hold-up in the oxygen and hydrogen separator tanks using mass and energy balance. The separators are modelled as ideal separators, hence the liquid/gas separation is perfect and all outflows are pure.
We hereby report the equations for the hydrogen separator (Fig. \ref{fig:separator}).
\begin{subequations}
\begin{align} 
    \dot n_{sep,2} & = f^{cat}_{out} - f_{sep,2}, \\
    \dot U_{sep,2} & = \tilde H_{in} - \tilde H_{out} \nonumber \\
    & = H(T, P_{sep,2},f^{cat}_{out}) - H(T_{sep,2}, P_{sep,2},f_{sep,2}).
\end{align}
\end{subequations}
The tank is assumed to be adiabatic.
Notice that the first equation is vectorial and that the outlet flow is 
\begin{equation}
    f_{sep,2}= \begin{bmatrix}
        f^{sep,2}_{H_2O} \\ f^{sep,2}_{H_2} \\ 0
    \end{bmatrix}.
\end{equation}
and its components are manipulated variables. 
It is necessary that the separator tanks hold a pressure that is never higher than the pressure in the electrolyzer $P$, in order for the flow not to reverse. Otherwise, a compressor may be used. Note that this is a control problem and it is part of the control structure of the plant.
The algebraic equations for the separator are
\begin{equation}
    g_{sep,1} = \begin{bmatrix}
        U_{TL} - U_{sep,2} \\ V_{TL} - V_{tot,2}
    \end{bmatrix}  = 0.
\end{equation}
The energy and volume evaluated by the thermodynamic library are
\begin{subequations}
    \begin{align}
    U_{TL} = & U(T_{sep,2},P,n^{sep,2,}_{H_2O})(l) + U(T_{sep,2},P,n^{sep,2}_{H_2})(g), \\
    V_{TL} = & V(T_{sep,2},P,n^{sep,2}_{H_2O})(l) + V(T_{sep,2},P,n^{sep,2}_{H_2})(g).
    \end{align}
\end{subequations}
The model for the gas-liquid separator for the oxygen stream is exactly the same, and we distinguish its variables by using the index 1.

\subsection{Compressor}
\begin{figure}[tb]
     \centering
     \includegraphics[width=0.27\textwidth]{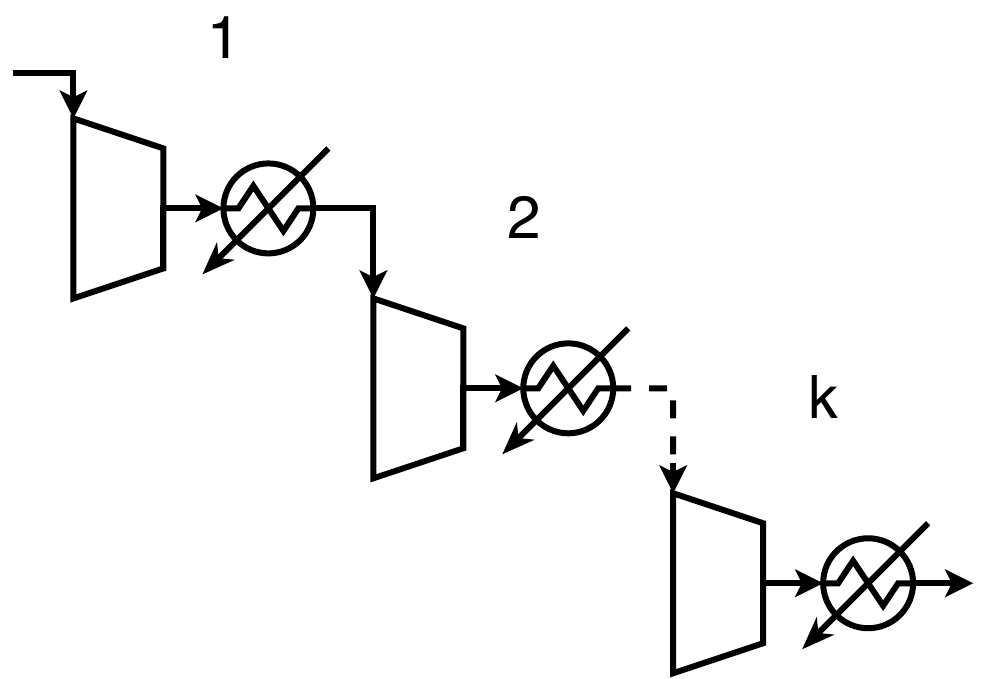}
     \caption{Multi-stage compressor with heat exchangers.}
     \label{fig:compressor}
 \end{figure}
 
The hydrogen that flows out of the separator needs to be compressed, before reaching the storage tank. Since the pressure requirement in such a tank can be very high, a multi-stage compressor is required. Figure \ref{fig:compressor} shows a $k$-stage compressor with cooling elements. Notice that determining the right amount of compressors to be used is a sizing problem. We choose 3 identical compressors. 
We model each compressor as isentropic (ideal and static), that means that the entropy stays constant during the compression process. For each stage $k$, we get an algebraic equation
\begin{equation}
    g_{k,comp} = \tilde S_{k,out} - \tilde S_{k,in} = 0.
\end{equation}
The entrophy flows are such that 
\begin{subequations}
    \begin{align}
        \tilde S_{k,in} & = S(T_{k,in},P_{k,in},f_{H_2}^{sep,2}), \\
        \tilde S_{k,out} & = S(T_{k,comp},P_{k,out},f_{H_2}^{sep,2}).
    \end{align}
\end{subequations}
$T_{k,in},P_{k,in}$ and $T_{k,comp},P_{k,out}$ are the temperature and pressure at the inlet and the outlet, respectively, of each compressor stage $k$.
The enthalphy change 
\begin{equation}
     \tilde H_{k,out} = \tilde H_{k,in} + \eta W_{comp,k} 
\end{equation}
allows us to compute the work $W_{comp,k}$ that every compressor has to perform.
$\eta$ is the efficiency of the compressor. The enthalpy flows are such that 
\begin{subequations}
    \begin{align}
        \tilde H_{k,in} & = H(T_{k,in},P_{k,in},f_{H_2}^{sep,2}), \\
        \tilde H_{k,out} & = H(T_{k,out},P_{k,out},f_{H_2}^{sep,2}).
    \end{align}
\end{subequations}
The intermediate cooling work by the heat exchangers between the stages can be computed, for each stage, as
\begin{equation}
    W_{hex,k} = \Delta H_{hex,k} = \tilde H_{k+1,in} - \tilde H_{k,out}.
\end{equation}
The last heat exchanger cools down to the tank temperature, therefore
\begin{equation}
    \tilde H_{4,in} = H(T_{tank}, P_{tank},f_{H_2}^{sep,2}).
\end{equation}

\subsection{Storage tank}
The mass and energy balance in the tank give the following differential equations
\begin{subequations}
\begin{alignat}{3}
\dot n_{tank} =& f^{sep,2}_{H_2} - f_{H_2}^{tank}, \\
\dot U_{tank} =& \tilde H_{in} - \tilde H_{out} - Q_{amb} = H(T_{tank},P_{tank},f^{sep,2}_{H_2}) - \\ &H(T_{tank},P_{tank},f_{H_2}^{tank}) - h_{tank} A_{tank} ( T_{tank} - T_{amb} ). \nonumber
\end{alignat}
\end{subequations}
$h_{tank}$ is the heat transfer coefficient of the tank and $A_{tank}$ is the transfer area.
 The following extra algebraic equations are necessary to solve the system
\begin{equation}
    g_{tank} = \begin{bmatrix}
        U_{TL} - U_{tank} \\ V_{TL} - V_{tank}
    \end{bmatrix}  = 0.
\end{equation}
The energy and volume evaluated by the thermodynamic library are
\begin{subequations}
    \begin{align}
    U_{TL} & = U(T_{tank},P_{tank},n_{tank}),\\
    V_{TL} & = V(T_{tank},P_{tank},n_{tank}).
    \end{align}
\end{subequations}

\subsection{Heat exchangers}
The purpose of the heat exchangers is to remove excess heat, produced by the reaction, from the water that is being recirculated back to the electrolyzer, hence the water outflows of the two separators.

We model the 2 heat exchangers with a static energy balance. The temperatures of the flows after the heat exchangers $T_{HX,1}$ and $T_{HX,2}$ are such that
\begin{equation}
    g_{HX} = \begin{bmatrix}
     \begin{matrix}
         H(T_{sep,1},P,f^{sep,1}_{H_2O}) - \tilde H_{HX,1} - Q_{HX,1}
     \end{matrix} \\ \begin{matrix}
          H(T_{sep,2},P,f^{sep,2}_{H_2O}) - \tilde H_{HX,2} - Q_{HX,2}
     \end{matrix}
\end{bmatrix}  = 0.
\end{equation}
$Q_{HX,1}$ and $Q_{HX,2}$ are the heat flows to be removed and are considered as manipulated variables. The outlet enthalphy flows are 
\begin{subequations}
    \begin{align}
        \tilde H_{HX,1} = &H(T_{HX,1},P,f^{sep,1}_{H_2O})  \\
    \tilde H_{HX,2} = & H( T_{HX,2},P, f^{sep,2}_{H_2O})
    \end{align}
\end{subequations}
\subsection{Flow mixer}
The mixing point before the electrolyzer stack inlet is modelled using static mass and energy balance. The following algebraic equations arise
\begin{equation}
    g_{mixer} =  \begin{bmatrix}
    f_{in,H_2 O} - f_{make-up} - f^{sep,1}_{H_2O} - f^{sep,2}_{H_2O} \\    
    \begin{matrix}
        \hfill H(T_{in}, P, f_{in,H_2O}) - \tilde H_{HX,1}  - \tilde H_{HX,2} \\ \hfill - H( T_{make-up},P, f_{make-up})
    \end{matrix}
    \end{bmatrix} = 0.
\end{equation}

\subsection{Full DAE system}
The full DAE system consists of 24 equations.
The differential variables are
\begin{equation}
\begin{split}
    x = [T,n^{sep,1}_{H_2O},n^{sep,1}_{O_2},U_{sep,1},n^{sep,2}_{H_20},n^{sep,2}_{H_2},\\ U_{sep,2},n_{tank},U_{tank}]^T.
\end{split}
\end{equation}
The algebraic variables are
\begin{equation}
    \begin{split}
    y = [ & \xi_{cell}, I,T_{sep,1},P_{sep,1},T_{sep,2},P_{sep,2}, T_{1,comp},T_{2,comp}, \\
     & T_{3,comp}, T_{tank},P_{tank}, T_{HX,1},T_{HX,2},f_{in, H_2 O},T_{in}]^T.
    \end{split}
\end{equation}
The manipulated variables (inputs) are
\begin{equation}
\begin{split}
    u = [f^{sep,1}_{H_2O},
        f^{sep,2}_{H_2O}, 
        f_{make-up}, f_{H_2}^{tank},
        Q_{HX,1}, 
        Q_{HX,2}, \\
        f^{sep,1}_{O_2},
        f^{sep,2}_{H_2}]^T.
\end{split}
\end{equation}
The disturbances are 
\begin{equation}
    d = [T_{amb}, P_{in}, T_{make-up}]^T.
\end{equation}
The output of interest of the system is
\begin{equation}
    z = f_{out,H_2}.
\end{equation}
\section{Simulation results} \label{sec:simulation}
\begin{figure}[tb]
    \centering
    \includegraphics[width=0.48\textwidth]{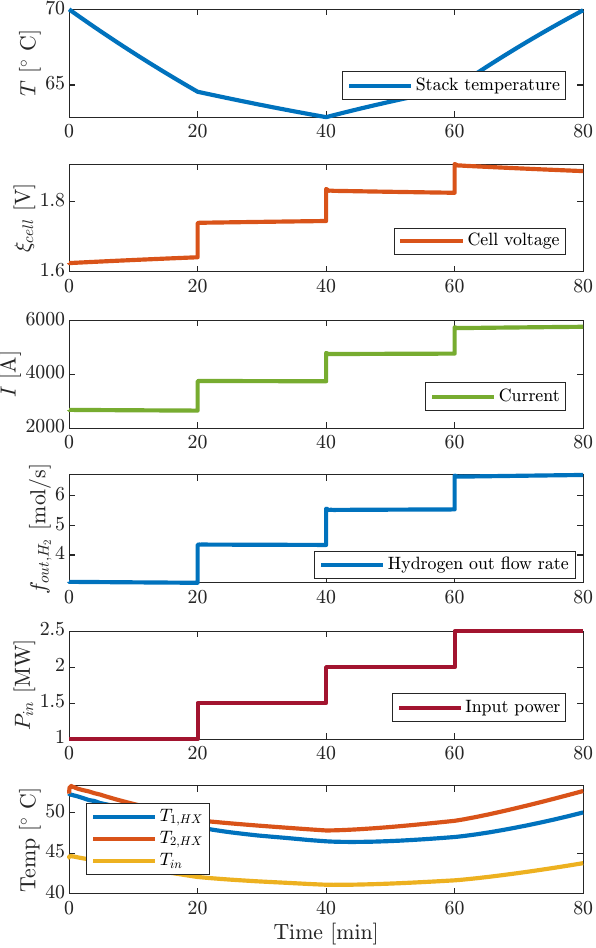}
    \caption{Dynamic simulation of the electrolyzer plant using step power input.}
    \label{fig:el_stack_simulation}
\end{figure}

We use the model parameters from \cite{RIZWAN202137120} for the electrolyzer stack. Table \ref{tab:paramters} reports them. 
We run a dynamic simulation of the model with step power input $P_{in}$ from 1 to 2.5 MW. The electrolyzer is assumed unpressurised ($P=1$ atm). The other disturbances are set to $T_{amb}=25 {}^\circ$ C and $T_{make-up}=30{}^\circ$ C. The inputs are kept fixed for the whole simulation to
\begin{equation}
\begin{split}
    u = [1 \text{ kg/s},1 \text{ kg/s},1 \text{ kg/s},2 \text{ mol/s}, 80 \text{ MW}, 80 \text{ MW}, \\
    1 \text{ mol/s}, \text{ mol/s} ]^T
\end{split}
\end{equation}
Fig. \ref{fig:el_stack_simulation} shows some of the most relevant variables from the simulation. These are the operating temperature of the electrolyzer, the cell voltage, the current, the hydrogen production rate and the power input. The temperatures after the heat exchangers and the mixer are also plotted.
The sudden change in power is reflected immediately in the hydrogen production rate, as the reaction rate directly depends on the current. 
The reader should note that even thought the relationship between input power and hydrogen production rate can be easily approximated as linear, the electrochemical dynamics and effect of having a water recirculation system with cooling make the overall system dynamics non-trivial. 
\begin{table}[tb]
    \centering
    \caption{Electrochemical model parameters \cite{RIZWAN202137120}.}
    \begin{tabular}{|c|l l|}
        \hline
        \textbf{Model parameter} & \textbf{Value} & \textbf{Unit} \\
        \hline

         $r_1$ & 2.18e-4 & $\Omega\,\text{m}^2$ \\
        $r_2$ & -4.25e-7 & $\Omega\,\text{m}^2 \, {}^\circ\text{C}^{-1}$ \\
        $s$ & 117.93e-3 & $\text{V}$ \\
        $t_1$ & -145.29e-3   &$ \text{m}^2\, \text{A}^{-1}$ \\
        $t_2$ & 11.794 & $\text{m}^2\, {}^\circ \text{C} \,\text{A}^{-1}$ \\
        $t_3$ & 395.68 & $\text{m}^2\,{}^\circ \text{C}^2 \,\text{A}^{-1}$ \\
        \hline
        $f_1$ & 120 & $\text{mA}^2 \text{cm}^{-4}$  \\
        $f_2$ & 0.98 & -  \\
        \hline
    \end{tabular}
    \label{tab:paramters}
\end{table}


\section{Conclusion} \label{sec:conclusion}
In this paper, we present a dynamic model for an alkaline electrolyzer plant. Each part of the plant is modeled using mass and energy balance. Thermodynamic properties (enthalpy, volume and internal energy) of the liquid and gases are evaluated using the thermodynamic library ThermoLib. We formulate the model as system of DAEs. Finally, we perform a dynamic simulation with variable input power. 
The results of the simulation confirm that the plant-wide dynamics are non-trivial, making the model a key enabler to process simulation and optimization. Future work include developing different controllers (PID and MPC) for the plant and using this simulation model to assess their performance.

\bibliographystyle{ieeetr}
\bibliography{Bibliography}

\begin{thebibliography}{10}

\bibitem{OLIVIER2017280}
P.~Olivier, C.~Bourasseau, and P.~B. Bouamama, ``Low-temperature electrolysis
  system modelling: A review,'' {\em Renewable and Sustainable Energy Reviews},
  vol.~78, pp.~280--300, 2017.

\bibitem{DAVID2019392}
M.~David, C.~Ocampo-Martínez, and R.~Sánchez-Peña, ``Advances in alkaline
  water electrolyzers: A review,'' {\em Journal of Energy Storage}, vol.~23,
  pp.~392--403, 2019.

\bibitem{Samani2020}
A.~E. Samani, A.~D'Amicis, J.~D. De~Kooning, D.~Bozalakov, P.~Silva, and
  L.~Vandevelde, ``Grid balancing with a large-scale electrolyser providing
  primary reserve,'' {\em IET Renewable Power Generation}, vol.~14, no.~16,
  pp.~3070--3078, 2020.

\bibitem{johnsen2023value}
A.~{Gloppen Johnsen}, L.~{Mitridati}, D.~{Zarrilli}, and J.~{Kazempour}, ``{The
  Value of Ancillary Services for Electrolyzers},'' {\em arXiv e-prints},
  p.~arXiv:2310.04321, Oct. 2023.

\bibitem{ULLEBERG200321}
Øystein Ulleberg, ``Modeling of advanced alkaline electrolyzers: a system
  simulation approach,'' {\em International Journal of Hydrogen Energy},
  vol.~28, no.~1, pp.~21--33, 2003.

\bibitem{SAKAS20224328}
G.~Sakas, A.~Ibáñez-Rioja, V.~Ruuskanen, A.~Kosonen, J.~Ahola, and
  O.~Bergmann, ``Dynamic energy and mass balance model for an industrial
  alkaline water electrolyzer plant process,'' {\em International Journal of
  Hydrogen Energy}, vol.~47, no.~7, pp.~4328--4345, 2022.

\bibitem{RIZWAN202137120}
M.~Rizwan, V.~Alstad, and J.~Jäschke, ``Design considerations for industrial
  water electrolyzer plants,'' {\em International Journal of Hydrogen Energy},
  vol.~46, no.~75, pp.~37120--37136, 2021.
\newblock International Symposium on Sustainable Hydrogen 2019.

\bibitem{HUANG202316}
C.~Huang, X.~Jin, Y.~Zong, S.~You, C.~Træholt, and Y.~Zheng, ``Operational
  flexibility analysis of alkaline electrolyzers integrated with a
  temperature-stabilizing control,'' {\em Energy Reports}, vol.~9, pp.~16--20,
  2023.

\bibitem{QI2023120551}
R.~Qi, J.~Li, J.~Lin, Y.~Song, J.~Wang, Q.~Cui, Y.~Qiu, M.~Tang, and J.~Wang,
  ``Thermal modeling and controller design of an alkaline electrolysis system
  under dynamic operating conditions,'' {\em Applied Energy}, vol.~332,
  p.~120551, 2023.

\bibitem{RITSCHEL20173542}
T.~K. Ritschel, J.~Gaspar, and J.~B. Jørgensen, ``A thermodynamic library for
  simulation and optimization of dynamic processes.,'' {\em IFAC-PapersOnLine},
  vol.~50, no.~1, pp.~3542--3547, 2017.
\newblock 20th IFAC World Congress.

\end{thebibliography}

\end{document}